\documentclass{desyproc}

\begin{document}
\title{The case for preserving our knowledge and data in physics experiments}

\author{{\slshape Frank Berghaus}\\[1ex]
University of Victoria, Victoria, Canada}

\contribID{Berghaus\_Frank}

\confID{13889}  
\desyproc{DESY-PROC-2017-XX}
\acronym{Patras 2017} 
\doi  

\maketitle

\begin{abstract}
  This proceeding covers tools and technologies at our disposal for scientific data preservation and shows that this extends the scientific reach of our experiments. It is cost-efficient to warehouse data from completed experiments on the tape archives of our national and international laboratories. These subject-specific data stores also offer the technologies to capture and archive knowledge about experiments in the form of technical notes, electronic logs, websites, etc. Furthermore, it is possible to archive our source code and computing environments. The paper illustrates these challenges with experience from preserving the LEP data for the long term.
\end{abstract}

\section{Introduction}
Data from particle physics experiments are collected with significant financial and human effort and are mostly unique. Experiment data are maintained by personal effort of scientists until they are judged no longer worth that effort. The tools and knowledge to interpret the data are lost as the members of the experiment move on to other projects. When later discoveries or new hypotheses for beyond-the-Standard-Model Physics renew interest in this forgotten data, theorists resort to recasting previously published results. Often this requires either extrapolations and assumptions or the need to be overly conservative in the possible interpretations because the original data or input needed to understand the data have been lost. This imposes limits on the scientific reach of our data. This is particularly worrisome as new physics is suspected in an increasing number of possible directions.

The physics programs of large experiments are completed only a few years upon the shutdown of the experiment. Without additional care, the utility of data diminishes even over the period of a few years as software and computing environments evolve and expert personnel and knowledge transition to new projects.

\section{Requirements For Long-Term Preservation}
The DPHEP Study group was initiated in 2009\footnote{As a subgroup of the International Committee on Future Accelerators} to ``to get a common vision on these issues and create a multi-experiment dynamics for further reference''. The study group identified the following tasks, in order of priority~:
\begin{enumerate}
  \item \emph{Experiment level projects}: The curation and preparation of data to be preserved must come from experts on the individual experiments. Some effort during the data-taking leads to a significant improvement in the ability to move experiment data to long-term preservation.
  \item \emph{International organization}: The responsibility of data preservation past the lifetime of an experiment must fall on a long-lived organization trusted by the experiment. The DPHEP collaboration~\cite{Amerio:2015ipk} was formed to coordinate that long-lived institutional support.
  \item \emph{Common R\&D projects}: Most challenges of providing long-term data preservation are shared between experiments. Concrete inter-experiment projects are facilitated by the DPHEP organization.
\end{enumerate}

\subsection{Knowledge, Data, and Software}
\label{Sec:kds}
To evaluate a new physics model against existing data one needs to run the detector simulation on the predictions of the physics model. The output of the detector simulation is reconstructed by the experiment reconstruction software, which yields a data format suitable for physics analysis. The detector simulation and reconstruction software are complex and require knowledge specific to each experiment. This knowledge is captured partially in internal documents during the operation of the experiment.

The software used by physics experiments is usually developed assuming a homogeneous computing environment and are therefore not portable. The large data volumes involved necessitate custom formats and protocols to optimize the performance of data acquisition, processing and analysis. Upon completion of the experimental programme the software and the ecosystem of supporting services are no longer maintained. Additional care needs to be taken when planning to access data in the long term and using experiment software given the pace of computing and software evolution.

Bit preservation of the scientific datasets as well as the knowledge, software, and computing environments identified above must be guaranteed. It is therefore advisable to choose an institution which operates the tools and services outlined in section~\ref{Sec:servicesandtools} as a trusted repository. There exist multiple international agencies that assess the trustworthiness of a repository for long-term preservation, for example the Data Seal of Approval\footnote{\url{https://datasealofapproval.org}}, the German nestor~\cite{JoDI199}, and the ISO 16363.

\section{Services and Tools}
\label{Sec:servicesandtools}
This section summarizes the tools and services commonly used to solve the challenges of data preservation. When evaluating the tools to preserve data one should keep in mind the target audience\footnote{Example target audience are the collaboration, fellow physicists, or the general public.} and timescale for preservation. These should be clearly stated in the data preservation policy and sets the effort and tools required.

\subsection{Digital Libraries}
Digital libraries provide databases to warehouse the documents that provide knowledge preservation. Care should be taken in curating the documents that are submitted to the digital library and a librarian should be considered to help to categorize these documents. The knowledge necessary to understand the experiment should not only be captured on documents held in the digital library but the target audience must be able to find these documents.

There exist many digital libraries, and some care should be taken when choosing the digital library to warehouse your documents. Generally, there are institutional (e.g.\ the CERN Document Server), subject-specific (HEPData), and generic digital libraries (Zenodo)\footnote{All of these are based on Invenio \url{http://invenio-software.org}}.

\subsection{Virtualization and Containerization}
Digital libraries are a suitable place to house the source code. Active maintenance of that source code usually ceases after the lifetime of an experiment. Software and hardware that source code relies upon continue to evolve. That means, on its own, the source code will eventually only serve as documentation. To use old binaries or compile the old source code the operating system, compiler, and necessary libraries need to be captured. Two technologies exist to facilitate this capture: virtualization and containerization. A virtual machine runs a complete operating system with drivers for hardware interfaces emulated by a host computer. A container is a user-space isolated by operating system features of the host. Virtual machines provide a mechanism of abstracting hardware architecture while containers do not. The choice should be informed by the period for which the data should be preserved.

Even virtual machines will be subject to some effects of software and hardware evolution. To ensure long-term preservation the virtual machine itself must be maintained. This maintenance is independent of the software contained in the virtual machine. Defining a test of the software environment such as a representative analysis that can be run by an automated system is advisable. An elegant solution to this problem is provided by the CernVM~\cite{Larsen:2012uy}. Note that containers may (and often are) run in virtual machines. The two technologies may readily be combined. Umbrella~\cite{Meng:2015} is a tool to determine and capture the dependencies of your software in a container environment.

\subsection{Tape Archives}
\begin{figure}
  \centering
  \includegraphics[width=0.75\textwidth]{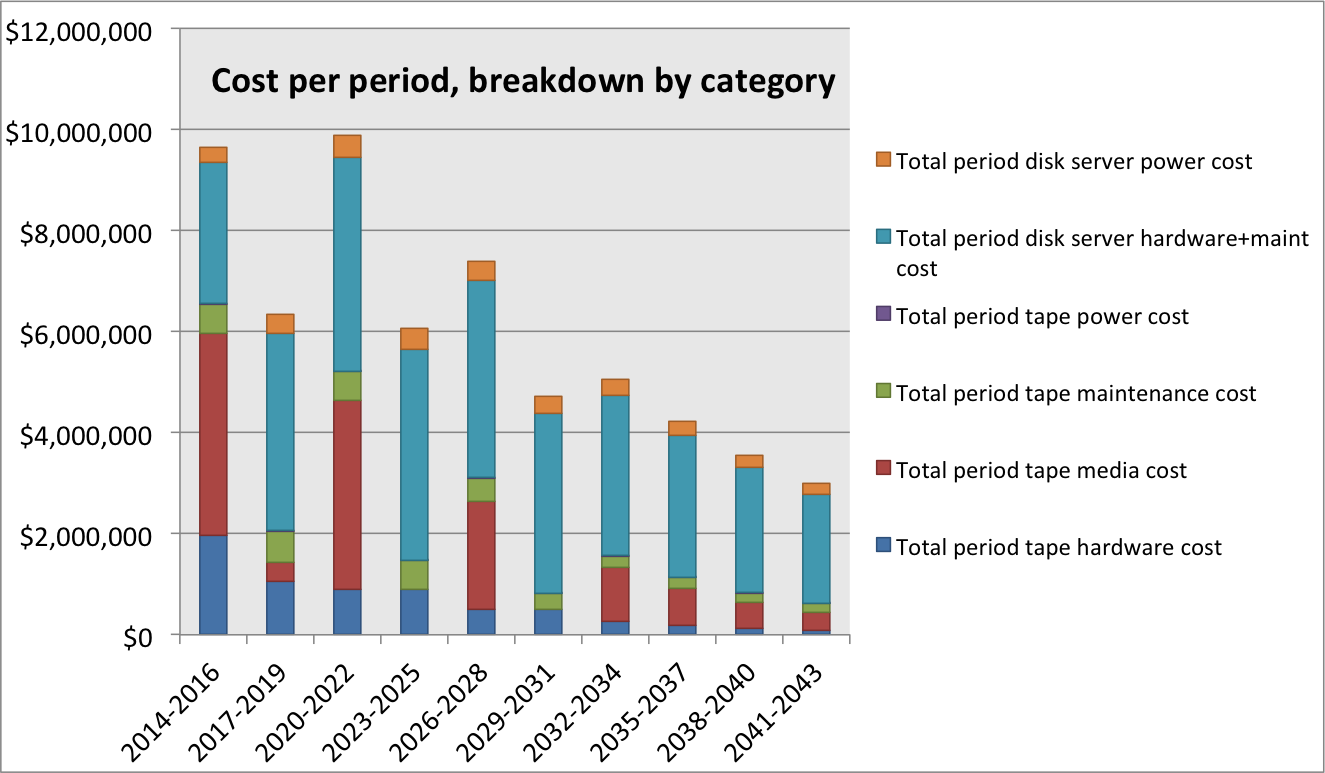}
  \caption{Cost projection for LHC Data on tape. The corresponding approximate growth in data volume is an increase from $\sim160$~PB in the 2014-2016 period to $\sim17$~EB in 2041-2043~\cite{Amerio:2015ipk}.}
  \label{Fig:tapecost}
\end{figure}
Documents in a digital archive, virtual machines, containers and scientific data are digital objects. Their long-term preservation is ensured by storage technology. Tape libraries are the current state of the art for rarely accessed data since they provide longevity\footnote{Tapes usually claim to provide data integrity for a few decades. To ensure data integrity in a large archive tape media is exchanged every 3-5 years.} at low cost. Figure~\ref{Fig:tapecost} shows a cost model for an LHC-sized data archive including purchasing and maintenance costs for tape libraries, drives and media\footnote{But excludes the cost of software development, licensing and the manpower for operation.}. The cost decreases even with the increasing data volume because the high cost of the tape libraries, robots, and readers are at the beginning while the additional tapes are relatively cheap. Furthermore the data density of the tape media is projected to increase by $\sim32$\% per year~\cite{Amerio:2015ipk}.

In the software design for the experiment, it is advisable to use common standards to access the data such as a POSIX filesystem or HTTP. While experiment or subject-specific tools provide high performance during active operation, support for the necessary protocols declines as the next generation of experiments adopts other tools and standards. Migration to those new standards is difficult.

\section{Case Study: LEP}
ALEPH, DELPHI, L3 and OPAL were the four experiments at the Large Electron-Positron Collider at CERN. Operation of the accelerator and experiments concluded in 2000. Research based on the data from these experiments is published to date, for example, search for the MSSM Higgs~\cite{Schael:2010aw} or a di-muon signal~\cite{Heister:2016stz}. The first shows the promise of data preservation, the second highlights the challenges of guaranteeing correctness of these analyses.

Before dissolving each LEP collaboration agreed on a data preservation and access policy. ALEPH chose to allow open access while the three others allowed analysis and publication only under the oversight of former members of the collaboration. As a result, ALEPH data may be found on open data catalogues\footnote{EUDAT at \url{http://b2find.eudat.eu/group/about/aleph}}, while those of the other three experiments may not. Revision of these policies proved difficult since the responsible members have moved on or think data access should remain restricted. Published and internal (technical) documents are captured by the CERN Document Server. The documentation of the experiments, including snapshots of their websites, publications and internal notes have been archived. Former collaborators from ALEPH, DELPHI, and OPAL built a running computing environment for their respective experiment~\cite{alephvm,delphivm,opalvm}. The results are virtual machine environments which provide access to the data of the experiments. The experiment software is provided using the CERN virtual machine file system. Long-term support of this file system was pledged by the LHC community.

The LEP experiments relied on RFIO to access data directly on tape, however, they all made access to data directly on the filesystem possible. Since support for RFIO has been discontinued access to the data is provided by a remote service\footnote{EOS Open Storage \url{http://eos.web.cern.ch} mounted using the XROOT fuse module.} mounted on the virtual machine. Access is authorized through the Kerberos authentication system at CERN. The data of the ALEPH collaboration is open. To access the data of OPAL and DELPHI a collaboration representative must be contacted\footnote{Contact information may be found at \url{https://greybook.cern.ch} or \url{http://dphep.org}}.

\section{Conclusions}
Criteria for planning and executing long-term data preservation strategies have been established. These strategies were tested with the LEP experiments to show that data analysis may be performed 15 years after last data taking. The software design decisions made during the development and operation of the experiment as well as the policies for data access have profound implications on the long-term usability and accessibility of scientific data. These criteria should be considered when planning new experiments.

\section*{Acknowledgements}
The author would like to thank J.~Shiers and the DPHEP collaboration who made this project possible. The work with the LEP experiments could only be accomplished thanks to the work of G.~Ganis and M.~Maggi (ALEPH), M.~Schr\"oder (OPAL), and U.~Schwickerath (DELPHI). The author would also like to acknowledge the many people from the CERN IT and libraries departments who continue to contribute to this work.


\begin{footnotesize}

\end{footnotesize}


\end{document}